# Performance Evaluation of Supervised Machine Learning Techniques for Efficient Detection of Emotions from Online Content

Muhammad Zubair Asghar[1], Fazli Subhan[2], Muhammad Imran[1], Fazal Masud Kundi[1], Shahboddin Shamshirband[3,4], Amir Mosavi[5,6], Peter Csiba[7], Annamaria R. Varkonyi-Koczy[5,7]

[1] ICIT, Gomal University, DIKhan and 29050, Pakistan; zubair@gu.edu.pk

[2] National University of Modern Languages, Islamabad, Pakistan; fsubhan@numl.edu.pk

[3] Department for Management of Science and Technology Development, Ton Duc Thang University, Ho Chi Minh City, Viet Nam; shahaboddin.shamshirband@tdt.edu.vn

[4] Faculty of Information Technology, Ton Duc Thang University, Ho Chi Minh City, Viet Nam

[5] Department of Automation, Kando Kalman Faculty of Electrical Engineering, Obuda University, Budapest, Hungary; amir.mosavi@kvk.uni-obuda.hu

[6] School of the Built Environment, Oxford Brookes University, Oxford OX3 0BP, UK; a.mosavi@brookes.ac.uk

[7] Department of Mathematics and Informatics, J. Selye University, Komarno 94501, Slovakia; csibap@ujs.sk & koczya@ujs.sk

**Abstract**: Emotion detection from the text is an important and challenging problem in text analytics. The opinion-mining experts are focusing on the development of emotion detection applications as they have received considerable attention of online community including users and business organization for collecting and interpreting public emotions. However, most of the existing works on emotion detection used less efficient machine learning classifiers with limited datasets, resulting in performance degradation. To overcome this issue, this work aims at the evaluation of the performance of different machine learning classifiers on a benchmark emotion dataset. The experimental results show the performance of different machine learning classifiers in terms of different evaluation metrics like precision, recall ad f-measure. Finally, a classifier with the best performance is recommended for the emotion classification.

# 1 Introduction

Cognitive science is defined as the interdisciplinary study of the mind. The emphasis of investigations in this domain is on the various human mental processes. These include sentiment, insight, thoughts, recollection, knowledge gaining, way of thinking, and emotions. Among them, emotion is deemed most significant in the area of human social behaviour identification. Over recent years, researchers have been looking into the employment of computational procedures for investigations on human emotions (Asghar et al. 2017a).

An emotion is a state of mind reflecting happiness, anger, disgust, fear, hate etc., and has a close association with human mood and feelings (Asghar et al. 2017b). Emotion detection from online content is relatively a new and challenging area in computational intelligence attracting attention of researchers in recent past.

Existing works (Asghar et al. 2017a, Asghar et al. 2017b, Sun et al. 2016, Jang et al. 2012, and Thomas et al. 2014) on the emotion-based sentiment classification systems are based on the lexicon-based and supervised machine learning (M.L) algorithms. The work performed by (Thomas et al. 2014) used a single machine learning classifier for the detection of emotion signals. However, we propose to apply five machine learning classifiers to detect seven categories of emotions. The proposed study is different from that of Thomas et al. (2014) in terms of increased number of machine learning algorithms and extended set of emotion signals (5 emotion signals).

## 1.1 Problem Statement

The emotion detection in public reviews is a challenging task due to its complex nature of emotion signals and their associated emotion words. The existing studies on emotion-based sentiment analysis using machine learning techniques (Thomas et al. 2014, Asghar et al. 2017, Sun et al. 2016) have used limited no. of classifiers and there is a lack of an extended combination of emotion signals for efficient classification of emotion in a given text. Therefore, it is required to develop an emotion-based sentiment analysis system using different machine learning classifiers for efficient classification of emotion express by the user in a given text by overcoming the limitations of the aforementioned studies. In this work, a supervised learning-based emotion analysis system is proposed with different machine learning classifiers for efficient emotion-based sentiment analysis.

## 1.2 Research Questions

RQ1. How to recognize and classify text-based emotions by applying M.L classifiers?

RQ2. What is the efficiency of different M.L classifiers with respect to different emotion signals?

RQ3: Which classifier is best for efficient emotion detection?

*1.3 Aims and objectives*

*1.3.1 Aim*

This work aims classifying emotions in a given text by applying multiple M.L algorithms and to suggest M.L algorithm with best classification results for the detection of different emotion signals

*1.3.2 Objectives*

1.To classify emotion in a given text using various supervised M.L algorithms by improving Thomas et al. (2014) work.

2.To evaluate the efficiency of different algorithms using different emotion signals.

3.To suggest a Machine Learning algorithm with high-performance results for emotion recognition.

*1.4 Research Significance*

The proposed system provides an application of various M.L algorithms in a given text. Second, the different emotion signals are applied to different machine learning classifiers (Guo et al. 2019, Anitescu et al. 2019), which are simple and effective. This would help computational intelligence experts in developing improved methods for the sentiment classification of text-based emotions.

The remainder of the article is outlined as follows: In section 2, related-work is presented; section 3 gives proposed method; results and discussion is described in section 4, and finally, section 5 outlines conclusion and future work.

**2 Related Work**

In this section, a review of the relevant studies is performed on emotion detection from online text

Thomas et al. (2014) proposed an emotion detection system which aims to classify sentences w.r.t different emotion classes. Experiments are conducted on ISEAR dataset using Naïve Bayes classifier. Different feature sets like uni-gram, bi-gram, and trigram, are applied using the weighted log-likelihood scoring technique. Promising results are achieved in terms of improved accuracy. However, experimentation with other classifiers is required.

Sun et al. (2016) proposed a cognitive model to interpret emotions from the complex text. The proposed system consists of four modules: (i) non-action centered, (ii) Metacognitive, (iii) action centered, and (iv) Motivational. An adoptive rule induction framework is proposed by identifying different emotion-related features. However, the performance of different algorithms is not evaluated with respect to their proposed system.

Emotions were extracted from different tweets using emotion-word hashtags and data set

"Hashtag Emotion Corpus" (Mohammad and Kiritchenko, 2015). A rich word-emotion dictionary was created using an emotion-labeled tweet dataset. Experimental results show that the SVM classifier performed better for basic emotion types. However, emotion words having different synonyms are not considered, which, if incorporated can improve the performance of the system.

Das and Yopadhyay (2012) proposed a sentence-level emotion detection system using Conditional Random Field and different lexicons, such as SenticNet, SentiWordNet (SWN) and WordNet affect. Additionally, the post-processing module along with emotion ranking technique is also proposed. Results show that their system achieved better performance as compared to the comparing method. The major limitation of their system is that it lacks comparison with supervised learning techniques.

Jang et al. (2012) worked on the development of emotion classification system using a machine learning algorithm. For this purpose, Support Vector Machine (SVM) and other algorithm are used for classifying emotion signals from the patient dataset. The SVM achieved the highest accuracy, however, system performance can be improved by performing experiments on different combinations of emotion signals.

Crossley et al. (2017) proposed a cognitive-based text analysis tool by implementing different text processing tasks including sentiment scoring using different lexicons to quantify user sentiments and emotion from different word vectors, are developed. However, the performance of the system can be improved by considering different variations of the n-gram features.

Cambria et al. (2012) proposed a Sentic computing-based technique for developing emotion analysis system by exploiting the rules of computer science and social science. Their technique works at the concept level and finds the context of the input text at a deeper level.

A sentence level emotion-based text analysis system is proposed by Shaila and Vadivel (2105) using a supervised learning technique. For this purpose, the Neural Network model is designed for isolating positive and negative emotions. It is reported that words and phrases have a significant role in emotion classification.

An automatic feedback analysis of student feedback is proposed by Kaewyong et al. (2015) using the lexicon-based technique. For this purpose, data acquisition is performed from more than 1100 student responses about teaching faculty. After applying different pre-processing techniques, opinion words are assigned sentiment scores using a sentiment lexicon. The proposed system shows improved results as compared to baseline methods.

An emotion detection system in E-learning domain is proposed by Binali et al. (2009). The system is capable of classifying student opinions regarding learning progress. Gate software is used to implement the framework.

Quan and Ren (2010) Proposed a polynomial Kernal technique based on a machine learning paradigm for calculating a similarity score between text and different emotion types. They achieved better performance with respect to the baseline method.

To detect emotion from facial expression in the video, Kollias et al. (2016) employed deep Convolutional Neural Network (DCNNs). The results show that the proposed method is effective with respect to comparing methods. However, the development of a real-life application for human-computer interaction can assist in evaluating the performance of the system more accurately.

A Chinese emotion lexicon is created by Li and Ren (2011) using Ren-CECPs (Corpus) for recognizing basic emotion types. An accuracy of 90% is achieved with respect to basic emotion types. However, performance can be improved further by extending the lexicon vocabulary.

To detect emotions from human speech, Davletcharova et al. (2016) implemented different speech recognition classifiers by employing various speech features, such as peak to peak distance. A dataset comprising of 30 different subjects was used, and better accuracy was achieved with respect to baseline methods.

Socher et al. (2013) proposed a deep learning module for classifying the sentences at a fine-grand level over a treebank corpus. For this purpose, the recursive Neural Network module is designed using training and testing data set. An accuracy of 80 to 85% is achieved as compared to the baseline method.

Jiang and Qi (2016) presented a chines emotion detection system for classifying user's emotions from online product reviews. For this purpose, an enhanced OCC-OR emotion model is used by selecting six emotion categories. The model is evaluated using different machine learning and natural learning techniques. The findings demonstrate the effectiveness of the proposed system.

Poria et al. (2016) proposed a convolutional learning technique for extracting emotions from multimedia content including audio video and text. An activation function is applied inside the inner layer and a performance improvement of above 80% was achieved with respect to comparing method.

Albornoz et al. (2012) proposed a concept-based emotion detection framework for classifying polarity for reputation. Different machine learning algorithms, including logistic regression and random forest, are used in Weka platform. However, the inclusion of subjectivity classification can improve the performance of the system.

Gao & Ai (2009) focused on the face gender classification using a multiethnic environment.in the literature AdaBoost was found very effective in accuracy and speed. Probabilistic bosting tree method was used. By experiment on snapshot and consumer images, PBT was found better than real AdaBoost methods.

Winarsih and Supriyanto(2016) evaluated the performance of different Machine Learning classifiers such as KNN, SVM, NB, and minimal optimization for emotion classification from Indonesian text. Different pre-processing steps such as tokenization, stop word removal stemming and case conversion are applied. Experiments are conducted using 10-fold cross-validation and result depict that the minimal optimization technology (SVM-SMO) performed better than the comparing methods.

Veenendaal et al. (2014) focused on the natural group emotion detection in indoor lighting. Emotional thinking has a side effect on memory and judgment. Edge detection was used with a Mesh superimposition to extract the features.

Rachman et al. (2016) developed an automatic emotion corpus using WordNet Effect for classifying emotions and affective norms for English words. Latent Dirichlet Allocation (LDA) technique is used for automatic expansion of the proposed corpus. Improved results are obtained with respect to comparing methods.

## 3 Material and Methodology

The proposed methodology includes the following models: (i) data collection, (ii) pre-processing, (iii) Applying different machine learning classifiers, and (iv) Comparison of different classifiers for emotion classification. Fig. 1 shows the overall working of the proposed system.

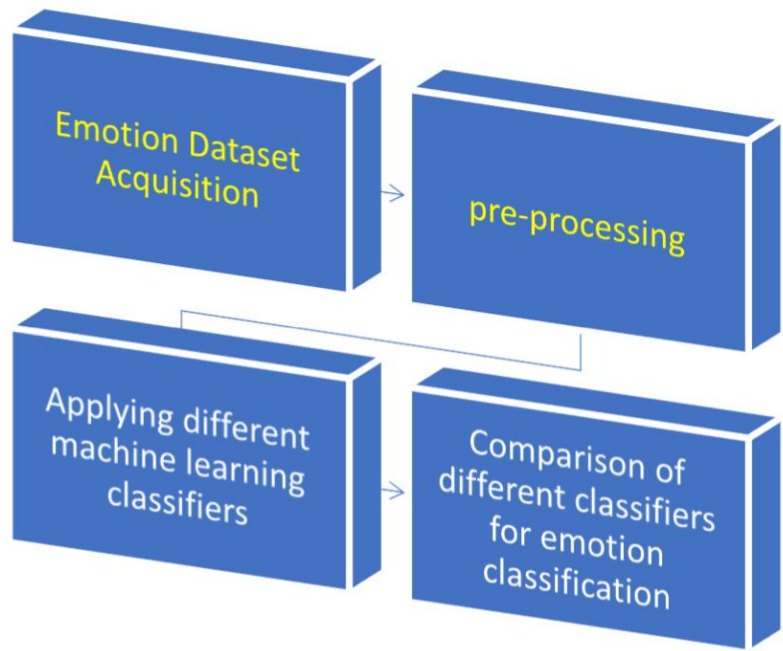

**Figure 1:** Diagram for Proposed System

### *3.1 Data collection*
A Publically available emotion-related data set, namely "ISEAR" (ISEAR, 2018), is used to conduct the experiments. The dataset is comprised of 2273 reviews. The sentences are annotated into Joy (1094), Fear (1095), Sadness (1096) Shame (1096), and Guilt (1096) emotions ("ideally divided into 5 classes"). We stored dataset into an MS Excel file and then converted into CSV files for conducting experiments. Table 1 shows the detail of the acquired dataset.

**Table 1:** Dataset Detail

| Title | Number of Reviews | Total Emotion Categories | No. of occurrences of each emotion category |
|---|---|---|---|
| ISEAR | 5477 | 5 (joy, fear, sadness, Shame, Guilt) | Joy(1094),fear(1095),sadness(1096),Shame(1096) Guilt(1096) |

## 3.2 Pre-processing

Different pre-processing tasks are carried out on the acquired dataset.

### 4.1.2 Tokenization

Tokenization breaks the sentence into small tokens using Python-based NLTK tokenizer.

### 4.1.3 Stop Word Removal

Different Stop words such as "a", "the", "am", etc. are eliminated using a predefined list implemented in python-based platform.

The pseudocode steps of the pre-processing module are presented in Algorithm 1 and the implementation code is shown in Appendix Table A.

**Algorithm 1:** Applying Pre-processing Step

```
PData← []
Dict← {}
row←0
while (row<=N-1)
        Words← Split (x [rows]," ")
        Preview ← []
        Repeat
                IF (final character of word = punctuation) {eliminate punctuation }
                IF (word is  not found in stop word list ){ Preview .Append (word)}
                IF (word is Not preent  in Dict){Dict[word]=0}

        Until (there is a word in Words)
        Append(Preview) to PData
end while
```

## 3.3 Applying Machine Learning Classifiers

In the next step, the input text is made an input to the different machine learning classifiers to get it classified into different emotion classes (joy, fear, sadness, shame, guilt). For this purpose, we implemented different supervised learning classifiers such as Naïve Bayesian (Danisman and Alpkocak, 2008), Decision Tree, KNN (K Nearest

Neighbour), Support Vector Machine, and Naïve Bayesian (Danisman and Alpkocak, 2008) using NLTK-based python framework (Loper and Bird, 2002).

*3.2.1 Feature Engineering*

To apply different machine learning algorithms, we used different feature selection steps, namely (i) counter vector creation, and (ii) tf x idf calculation.

***Count Vector creation:*** The count vector, also called vocabulary-of-words is a popular encoding scheme to a constitute word vector for a given document (Jason Brownlee, 2017).

 ***(TF-IDF): Term Frequency- Inverse document Frequency*** (TF-IDF) is an important feature representation scheme, where TF shows the frequency of a term in a given document and DF is the frequency of a given document in the total no. of documents. This measure is important because it shows the importance of a given term, instead of the traditional frequency count  *(*Patil, L. H., & Atique, M. 2013).

The pseudocode steps of the feature engineering module are presented in Algorithm 2 and the implementation code is shown in Appendix Table A.

**Algorithm 2:** Steps for Feature Engineering

```
# Count Vector
SET CVector to []
For     Each review in PData Do
            For each word in review DO
                Set Dict [word] to Dict [word] +1
            End For
            .Append Dict to Vector  Assign 0 to  Dict values
End For
# Term Frequency
SET TF to cvector
Row←0
While (row <= N-1) Do
        SET NOwords to SUM (Cvector [row].values)
        For Each W in Cvector [row]
        SET TF [W] to Cvector [W]/Nwords
        End For
While end
# TF/ DF
# IDF Computation
```

```
SET IDF TO []
Wile (there exist a row in TF) Do
Temp← { }
While (there exist a word in  in row ) DO
        Count← 0
        For i from 0 to N-1 Do
                IF TF [c][word]>0 then
                    Count← count+1
                        End IF
                End For
                Assign LOG (N/Count) to Temp [word]
        While end
        Append Temp to IDF
While end#TF-IDF
SET TF-IDF to []I←0
While (I<=N-1)
        Assign { } to  Temp
        For Each Word in TF [i], IDF [i]
                Temp [word]= TF[i][word]*IDF[i][word]
        End For
        .Append Temp to  TF-IDF
End While
```

*3.3.2 Splitting dataset into Train and Test*

The dataset acquired is dissected into training (80%) and testing (20%) chunks. A set of reviews with respect to training and testing sets are shown in Table 2 and Table 3 respectively.

**Table 2:** A sample Listing of Training Dataset

| Review No | Reviews | Reviews emotion |
|---|---|---|
| 01 | (33) My 2 year old son climbed up and sat on the 7th floor balcony with his legs hanging out.  He was holding on tightly to   the upper railing of the balcony but he could have easily lost   his balance when he sat down. | Fear |
| 02 | (35) I saw my 18 year old son grab an oxygen mask as he had breathing difficulties.  I had a bad conscience because I had | Shame |
|    | not stopped smoking. Medication for the dilation of his breathing tubes also caused a sense of guilt in me. | |
| 03 | (66) I complained about a colleague to the manager and he told her that someone had complained; this colleague came to me believing that I liked her. | Guilt |
| 04 | (49) I saw my 18 year old son grab an oxygen mask as he had breathing difficulties. I had a bad conscience because I had not stopped smoking. Medication for the dilation of his breathing tubes also caused a sense of guilt in me. | Shame |
| 05 | (67) At a friend's birthday party with some of my closest friends. It was all very pleasant and one could say that I was happy to have such good friends. | Joy |
| 06 | At my Summer job a new caretaker had been employed who was my age but I preferred going out for lunch with the accounts personnel rather than with him. | Guilt |
| 07 | (103) I saw the list of books to read for an accounting course, I thought \ Oh God how I will ever manage it! | Sadness |
| 09 | (77) The whole family gets together for a one week holiday. Everybody feels free and the trip is well planned. It works out well and we enjoy ourselves. | Joy |
| 10 | (127) I am dishonest to a friend to whom I am very close. I feel guilty because I know that he gives a different version of the truth and I have not corrected these mistakes, and he is aware that I know that they are wrong. | Guilt |
| 11 | (138) I did not get the salary increase that I had been expecting and understood how little one's work was appreciated. | Sadness |
| 13 | (148) Due to laziness, I failed the term studies completely at University. I also wanted, to some extent, to protest against my parents' expectations. | Guilt |
| 14 | (210)After having quarrelled unnecessarily and without any reason, and having been stupidly cross in every way. | Shame |
| 16 | (243) I made the same mistake that I had accused someone else of, and this was obvious to a third person. | Shame |
| 17 | (310) I was at the end-of-term party last week and had fun as happy and sang and drank only soft drinks. It is possible to have a good time without alcohol. | Joy |
| 18 | (412) After having slept for a short time I woke up - I had the feeling of someone standing beside me and was very frightened. I had to turn on the light turn on the light for several minutes before I was able to get to sleep again. | Fear |
| 19 | (242) I thought that it was going to be impossible for me to start studying (due to wrong information from the student advisor). | Sadness |
| 20 | (1090)At school I was bad in mathematics. Although my teacher had admonished me to do my homework, one day I had forgotten to do it. When my teacher noticed it, I blushed and was ashamed to be rebuked in front of the other pupils. | Shame |
| 21 | (2060)I had to have my tonsils out. I had been making up my mind almost for a year - I was afraid. But during the two hours in the hospital room, while I was waiting to be called for the operation I felt a real fear - of the pain, of what they were going to do to me, of the unknown. | Fear |
| 22 | (2375) I am ashamed at myself sometimes when I am working with handicapped people and don't wish to be seen in public with them. | Shame |
| 23 | (2189) I was admitted to the Institute. I had problems with many people about my applying for this institute and I decided to prove that one can pass excellently without visiting any preparatory course. That is - I proved what I could do on my own. | Joy |
| 24 | (2631)I felt very sad when I left home because I could not stand it any longer. I do not regret it, but I missed my little sister very much (and she missed me). These feelings wear off over time. | Sadness |
| 25 | (2905) My grandmother several times has been struck by cerebral haemorrhages. Until now she recovered well each time, but there is always the threat to lose her. | Fear |

| 26 | (3205) One night, I went out with some friends for dinner and I did not tell my parents that I would come back late. I thought of phoning but in the end I did not. When I arrived home, my parents were very worried. | Guilt |

**Table 3:** A sample Listing of Testing Dataset

| Reviews No | Reviews | Reviews Emotions |
|---|---|---|
| 08 | (3236) My brother was born, everything went all right. It had been very likely that he would have a deficiency (my sister suffers from Down's Syndrome) and that my mother would be in danger. | Joy |
| 12 | (311) I had a long discussion on politics with an acquaintance. He was more knowledgeable than me and I failed to explain my point of view and was misunderstood so I felt depressed and left. | Sadness |
| 15 | (387) We got lost in Florence and the coach did not turn up until midnight. I had no place to go to and there were strange reports at the police station. | Fear |
| 27 | (3251) The final marks were to be given in the morning. I wanted to get there late because I was very afraid. When I arrived there, everybody was very happy and I had also passed. We had a wonderful time all day long. | Joy |
| 28 | (4207) The teacher asked me a question in class, concerning something I had read earlier, and I did not know the answer, so I felt ashamed in front of the whole class. | Guilt |
| 29 | (4644) I once felt guilty when a certain passenger in the same (ship, plane?) which I boarded when coming from home lost his 10t which in fact fell into my pocket unknowingly. | Guilt |

The pseudocode steps of the splitting dataset into train and test, are presented in Algorithm.3 and the implementation code is shown in Appendix Table A.

**Algorithm 3:** Steps for Splitting dataset into Train and Test

```
# Dataset Splitting into train/Test
X_train ← []
Y_train ← []
X_test ← []
Y_test ← []
test-size ← 20% of N
Assign RANDOM(0,N-1,Test-size) to TIndices
I ← 0
While (i<=N-1) do
temp ← []
Repeat
    Append(IF-IDF[i][word]) to Temp
Until (there is word in TF-IDF[i])
        IF I exists in TIndices then
                Append Temp to X_test
                Append review_text[i][1] to Y_test
        Else
                Append Temp to X_train
                Append Temp to Y_train
        End IF
        I++
While end
```

In the following sub-sections, different machine learning classifiers used in this study, are summarized.

*3.3.3 Naïve Bayes*

The Naïve Bayes (N.B) machine learning technique is based on the Bayes theorem, belonging to a family of probabilistic classifiers (Liu B, 2002). The features and attributes used are self-reliant from each other, forming a naïve assumption. It is formulated as follows:

$$P(H|P) = \frac{P(h|p)P(h)}{P(p)} \quad (1)$$

*3.3.4 Random Forest*

Random Forest (RF) technique is one of the frequently applied ML algorithm, based on the findings acquired from decision tree generated during training (Liu B, 2002). The output of the forest is the focused output from each decision tree. Mathematical representation is presented as follows:

Let D = $\{(x_{1}y_{1}............x_{n}y_{n})\}$

Where $x_i$ is prediction and $y_i$ is target variable
h = {h_i(x)..............h_k (x)}
Where h is ensemble of classifier,
$h_k(x)$ is a decision tree
f(x) = f[{h_k (x)}]
Where x is input, and each tree cash a vote for the most popular class at input x, and the class with most votes wins.

### 3.3.5 Support Vector Machine
The Support Vector Machine (SVM) is based on the binary and multi-classification, classifying all the text into different emotion categories. In order to classify the text into different emotion classes (joy, fear, sadness, shame, guilt), SVM finds the maximum margin hyperplane, mathematically, it is formulated as follows:
Eq. 1 for an example. The number should be aligned to the right margin.

$$D=\{(t1,d1),(t2,d2),\ldots\ldots\ldots\ldots(tn,dn)\} \tag{2}$$

### 3.3.6 Logistic Regression
The Logistic Regression (LR) performs the classification of text into multiple emotion types using training and testing sets (Varathan et al. 2017). It is predicted that to which emotion class/tag, the text belongs (Algorithm 4). The LR is the fast prediction algorithm. Its mathematical formulation is presented as follows:
Eq. 1 for an example. The number should be aligned to the right margin.

D= {(X, y_i)} (Data)

$$(Prediction) = \frac{1}{1+e^{-(b0+bX)}} \tag{3}$$

Where $b_0$ is the bias (intercept), and X is the input vector, b is the coefficient of input.
Updating (coefficient value)
b= b+∝ × (y-prediction) × prediction × (1-prediction) ×X
Where ∝ is learning rate, Y target variable, X input prediction

$$P(Emotion\,|X) = \frac{1}{1+e^{-(bo+bx)}}$$

Output having maximum probability will be selected as prediction.

**Algorithm 4:** Steps for Logistic Regression classifier

---
*Train/ Test (Logistic Regression Classifier)*
Let X= Explanatory Variable (vector)
Let Bm,k = regression Coefficient associated with mth explanatory variable and kth emotion (outcome)
# Observation i and outcome k
Let P(k,i)=$B_k$ . $X_i$
For i For 1 to 4 Do
P(Emotion=ei)=$\frac{e^{B}\cdot X_i}{1+\sum e^{B}ei.Xi}$
Select e$i$ $as\ a\ outcome\ having\ maximum\ probability$
End for
#Train/ Test (SVM)

---

*3.3.7 K-Nearest Neighbor*

The K-Nearest Neighbor (KNN) model performs both classification and regression, based on instance-driven learning. In the emotion classification work, the KNN uses majority voting of its neighbors for tagging the text with particular emotion category. It is formulated as follows:

$$1D \sum ai \in M\ D(a)Yi \qquad (4)$$

*3.3.8 XG Boost*

Extreme Gradient Boosting (XGBoost) classifier is based on the gradient boosting framework (Babajide Mustapha & Saeed 2016). It assists in solving most of the prediction and classification problems in data science efficiently. The classifier is robust and yields efficient results in different distributed environments, such as Hadoop, SGE and MPI. Algorithm 5 shows working of XGboost classifier.

**Algorithm 5:** Steps for XGboost classifier working

Let D = {($x_i$, $y_i$)}$^n$ i=1 is training data
L (y,F(x)) is Loss function, M is number of iteration
F0(x) = argmin y $\sum_{i=1}^{n} L(yi, Y)$     [Initialize model….constant value]
For m=1 to M
$Y_{im}$ = -[$\frac{\hat{o}L(yi,F(xi))}{\hat{o}F(xi)}$] F(x)= $F_{m-1}$ (x)
X compute multiples $Y_m$
$Y_m$ = argmin $\sum_{i=1}^{n} L(yi, \text{Fm} - 1(xi) + Yhm(xi))$
Update the model
$F_m$ (x) = $F_{m-1}$ (x) + $Y_m h_m$ (x)
Output
Fm(x)

### 3.3. 9 The Complete Algorithm

The pseudocode steps of the proposed system, are presented in Algorithm 6 and the implementation code is shown in Appendix Table A.

**Algorithm 6:** Pseudocode of the Proposed System

| | |
|---|---|
| Input: | set of sentences in ISEAR dataset saved in excel workbook |
| Output: | Text classified into Emotion category |
| Emotion Category: | ["Joy", "Fear", "Sadness", "Shame", "Guilt"] |
| ML-Classifiers: | ["SVM", "NB", "KNN", "XGboost","SGD classifier", "Random forest", "Logistic regression"] |
| Stop-word List: | [this, that, is, was by…….] |

**Start**
*//Text Scanning*
Text ← Read text from dataset
*#Applying Pre-processing (tokenization/stop words removal/punctuations)*
*#Tokenization/segmentation*
    Token← tokenize (text)
*# Stop words elimination*
    P_text ←eliminate_ stop words (tokens)
*#punctuation*
*# data set splitting into train/test*
    Assign Split (p_text, test-size=20%) to X-train, Y-train, x-test, y-test
*# counter-Vector creating (p_text)*
#tf-idf computations
#applying classifier
Assign Classifier () to Model
Assign Model: fit(x-train, y-train) to Classification
#Prediction
Assign classification: Prediction (x-text) to Prediction
*#Accuracy*

```
  Assign Accuracy (Prediction, t-text) to Accuracy
  #Configuration Matrix
  Assig Confusion matrix (y-test, prediction) to CF
  #performance   evaluation using precision, recall, F-Measure
  Output_emotion ←classification –report (y-text, prediction, emotion_ category)
  Return (Output_emotion)
```

### 3.4 Comparison of different Classifiers for Emotion Classification

After applying the aforementioned classifiers for emotion detection in the text, we have applied different performance evaluation measures like Precision, Recall and F-measure (Quan & Ren, 2010).

The obtained results are presented in section 4 "Results and Discussion". The pseudocode steps of the performance evaluation of the different classifiers are presented in Algorithm 7 and the implementation code is shown in Appendix Table A.

**Algorithm 7:** Pseudo code of the Performance Evaluation

```
# Performance
 Assign count (Prediction =Y-test) TC
Assign TC/N2 to Accuracy←
Assign COUNT (Prediction = Positive AND Y_test=Paositive) to TP
Assign COUNT (Prediction =Negative AND Y_Test= Negative) to TN
Assign COUNT (Prediction=Positive and Y_test=Negative) to FP
Assign COUNT (Prediction=Negative and Y_test=Positive) to FN
Assign TP/(TP+FP) to Precision
Assign IP/(TP+FN) to Recall
Assign { } to FFM
Assign TP to CFM ['TP']
Assign FN to CFM['FN']
Assign FP to CFM['FP']
Assign TN to CFM['TN']
```

## 4 Results and Discussion

To evaluate the performance of the proposed system with respect to emotion classification, various evaluation measures including accuracy, precision, recall, and F measure are employed. In the rest of the sub-sections, we try to answer the posed research questions by analyzing the findings of the conducted experiments.

### 4.1 RQ1: How to recognize and classify text-based emotions by applying M.L classifier?

To answer this research question different supervised Machine learning classifier such as SVM, Random Forest, Naïve Bayesian, Logistics, KNN, XG boost, stochastic gradient, and BPN, are implemented using Python and Jupiter notebook (Ragan-Kelley et al. 2014). For this purpose, the acquired dataset is divided into training (80%) and testing

(20%) blocks. The basic aim of the aforementioned classifiers is to predict appropriate emotion labels, namely *Joy, Fear, Sadness, Shame, and Guilt.*

*4.1.1 Classifier with best Performance*

Results shown in Table 5 show that logistic regression performed well with respect to accuracy (avg) (66.58%), recall (avg) (0.67), and precision (avg)(67), as compared to other classifiers. As far as F1-score (avg) (66%) is concerned, both logistic regression, as well as SGD classifier, performed well.

*4.1.2 Classifier with worst performance*

Results presented in Table 6 show that the K Nearest Neighbour (KNN) produced lowest performance in terms of precision (avg) (0.58), recall (avg) (0.58), f1 score (avg) (0.57), and accuracy (avg) (57.81%).

*4.2 RQ2: What is the accuracy of different M.L classifiers with respect to different emotion signals?*

To answer this research question, we conducted a number of experiments to evaluate the performance of different supervised machine learning classifiers with respect to emotion classification. These experiments were conducted on a PC with an Intel Core i5-2450M processor with a 3.0-GHz clock speed. The times show the average amount of CPU time used to classify instances (750) in the dataset.

*4.2.1 Experiment#1*

Experiment#1 is conducted to evaluate the performance of Support Vector Machine (SVM) with respect to emotion classification. The performance evaluation results in emotion classification using SVM in Table 4. The results depict that the SVM classifier achieved the best performance with respect to F measure (77%) precision (76%) and Recall (77%) for "joy" emotion tag. The SVM classifier produced the best recall and F1-score results of (77%) for "Joy" emotion tag and overall accuracy of 64.66%. The CPU time (speed) of SVM classifier is also reported (2.43).

**Table 4:** Performance Evaluation Results using SVM

| Emotion Tags | Precision | Recall | F1-score |
|---|---|---|---|
| Joy | **0.76** | **0.77** | **0.77** |
| Fear | 0.54 | 0.62 | 0.58 |
| Sadness | 0.75 | 0.73 | 0.74 |
| Shame | 0.67 | 0.56 | 0.61 |
| Guilt | 0.54 | 0.55 | 0.55 |
| **Accuracy** | 64.66% | | |

| CPU Time (ms) | 2.43 |
|---|---|

*4.2.2 Experiment#2*

Experiment#2 is conducted to evaluate the performance of Logistics regression with respect to emotion classification. The performance evaluation results for emotion classification using Logistics regression, are presented in Table 5 the results depict that the Logistic regression classifier achieved the best performance with respect to F-measure (76%) for "joy" emotion tag, and recall (83%) for "Joy" emotion tag. Similarly, a precision of 73% is attained for "sadness" emotion tags and overall accuracy of 66.58%. The CPU time (speed) of Logistics Regression classifier is also reported (4.11).

**Table 5:** Performance Evaluation Results using Logistics Regression

| Emotion Tags | Precision | Recall | F1-score |
|---|---|---|---|
| Joy | 0.70 | **0.83** | **0.76** |
| Fear | 0.62 | 0.67 | 0.64 |
| Sadness | **0.73** | 0.73 | 0.73 |
| Shame | 0.70 | 0.55 | 0.62 |
| Guilt | 0.58 | 0.56 | 0.57 |
| **Accuracy** | 66.58% | | |
| **CPU Time (ms)** | 4.11 | | |

*4.2.3 Experiment#3*

Experiment#3 is conducted to evaluate the performance of KNN with respect to emotion classification. The performance evaluation results of emotion classification using KNN are shown in Table 6 The results depict that the KNN classifier achieved the best performance with respect to precision (66%), F-measure (67%) recall of 68% for *"Joy" and "sadness"* emotion tags and overall accuracy of 57.81%. The CPU time (speed) of KNN classifier is also reported (19.45).

**Table 6:** Performance Evaluation Results using KNN

| Emotion Tags | Precision | Recall | F1-score |
|---|---|---|---|
| Joy | **0.66** | **0.68** | **0.67** |
| Fear | 0.53 | 0.55 | 0.54 |
| Sadness | **0.66** | **0.68** | **0.67** |
| Shame | 0.49 | 0.65 | 0.56 |
| Guilt | 0.59 | 0.33 | 0.42 |
| **Accuracy** | 57.81% | | |
| **CPU Time (ms)** | 19.45 | | |

*4.2.4 Experiment#4*

Experiment#4 is conducted to evaluate the performance of Naïve Bayesian (N.B) with respect to emotion classification. The performance evaluation results for emotion classification using Naïve Bayesian (N.B), are shown in Table 7 The results depict that the Naïve Bayesian (N.B) classifier achieved the best performance with respect to precision (76%) for "sadness" category, F-measure (73%) for *"Joy"* category, and a recall of (0.75) for "Joy" emotion tag and overall accuracy of 63.6%. The CPU time (speed) of NB classifier is also reported (13.43).

**Table 7:** Performance Evaluation Results using Naïve Bayesian

| **Emotion Tags** | **Precision** | **Recall** | **F1-score** |
|---|---|---|---|
| Joy | 0.71 | **0.75** | **0.73** |
| Fear | 0.55 | 0.64 | 0.59 |
| Sadness | **0.76** | 0.65 | 0.70 |
| Shame | 0.61 | 0.60 | 0.60 |
| Guilt | 0.59 | 0.56 | 0.57 |
| **Accuracy** | 63.6% | | |
| **CPU Time(ms)** | 13.43 | | |

*4.2.5 Experiment#5*

Experiment#5 is conducted to evaluate the performance of Random forest RF (200) with respect to emotion classification. The performance evaluation results of emotion classification using Random forest RF are shown in Table 8 The results depict that the Random forest RF (200) classifier achieved the best performance with respect to precision (69%) for "sadness" emotion category, F measure (71%) and a recall of 76% for "Joy" emotion tags and overall accuracy of 64.02%. The CPU time (speed) of RF classifier is also reported (7.61).

**Table 8:** Performance Evaluation Results using Random forest RF (200)

| **Emotion Tags** | **Precision** | **Recall** | **F1-score** |
|---|---|---|---|
| Joy | 0.67 | **0.76** | **0.71** |
| Fear | 0.63 | 0.58 | 0.61 |
| Sadness | **0.69** | 0.70 | 0.70 |
| Shame | 0.68 | 0.56 | 0.61 |
| Guilt | 0.54 | 0.61 | 0.57 |
| **Accuracy** | 64.02% | | |
| **CPU Time (ms)** | 7.61 | | |

*4.2.6 Experiment#6*
Experiment#6 is conducted to evaluate the performance of the XG Boost (extreme gradient boosting) with respect to emotion classification. The performance evaluation results of emotion classification using XG Boost (extreme gradient boosting) are shown in Table 9. The results depict that the XG Boost (extreme gradient boosting) classifier achieved the best performance with respect to precision (66%) for "Joy", "sadness" and "shame" emotion tags, and F measure (66%) for "joy" and "sadness" emotion tags, whereas a recall of 66% is attained for "Joy" emotion tag and overall accuracy of 58.54%. The CPU time (speed) of XG Boost classifier is also reported (2.01).

**Table 9:** Performance Evaluation Results using XG Boost (extreme gradient)

| Emotion Tags | Precision | Recall | F1-score |
|---|---|---|---|
| Joy | **0.66** | **0.66** | **0.66** |
| Fear | 0.56 | 0.56 | 0.56 |
| Sadness | **0.66** | 0.65 | **0.66** |
| shame_ | **0.66** | 0.49 | 0.56 |
| Guilt | 0.44 | 0.56 | 0.50 |
| **Accuracy** | 58.54% | | |
| **CPU Time(ms)** | 2.01 | | |

*4.2.7 Experiment#7*
Experiment#7 is conducted to evaluate the performance of SGD Classifier (Stochastic gradient) with respect to emotion classification. The performance evaluation results of emotion classification using SGD Classifier (Stochastic gradient) are shown in Table 10. The results depict that the SGD Classifier (Stochastic gradient) classifier achieved the best performance with respect to precision (75%) for "sadness" emotion tag, F measure (75%) and recall of 77% is attained for "Joy" emotion tag and overall accuracy of 65.57%. The CPU time (speed) of SGD classifier is also reported (6.11).

**Table 10:** Performance Evaluation Results using SGD Classifier (Stochastic gradient)

| Emotion Tags | Precision | Recall | F1-score |
|---|---|---|---|
| Joy | 0.73 | **0.77** | **0.75** |
| Fear | 0.53 | 0.62 | 0.57 |
| Sadness | **0.75** | 0.74 | 0.74 |
| shame_ | 0.70 | 0.62 | 0.65 |
| Guilt | 0.60 | 0.54 | 0.57 |
| **Accuracy** | 65.57% | | |
| **CPU Time(ms)** | 6.11 | | |

*4.2.8 Experiment#8*
Experiment#8 is conducted to evaluate the performance of BPN Classifier (Back Propagation Neural) model with respect to emotion classification. The performance evaluation results of emotion classification using BPN Classifier are shown in Table 11. The results depict that the BPN Classifier achieved the best performance with respect to precision (72%) for "guilt" emotion tag, F measure (73%) and recall of 75% is attained for "Joy" emotion tag and overall accuracy of 71.27%. The CPU time (speed) of BPN classifier is also reported (3.29).

**Table 11:** Performance Evaluation Results using BPN (Back Propagation Neural Classifier)

| **Emotion Tags** | **Precision** | **Recall** | **F1-score** |
|---|---|---|---|
| Joy | 0.71 | **0.75** | **0.73** |
| Fear | 0.51 | 0.55 | 0.57 |
| Sadness | **0.71** | 0.68 | 0.69 |
| shame_ | 0.69 | 0.64 | 0.66 |
| Guilt | **0.72** | 0.66 | 0.69 |
| **Accuracy** | 71.27% | | |
| **CPU Time (ms)** | 3.29 | | |

*Over all Result table*
*4.3 RQ: 3 Which classifier is best for efficient emotion detection?*
This experiment aims at inspecting the results obtained from previous experiments and recommendation is made for the emotion detection classification on the basis of comparing results obtained from different classifiers (Table 12).

**Table 12:** Overall Results

| Classifier | Emotions | Precision (Avg) | Recall (Avg) | F1-Score (Avg) | Accuracy |
| --- | --- | --- | --- | --- | --- |
| XGBoost | Joy, Fear, Sadness, Shame Guilt | 0.60 | 0.59 | 0.59 | 58.54 |
| Support Vector Machine (SVM) | Joy, Fear, Sadness, Shame Guilt | 0.65 | 0.65 | 0.65 | 64.66 |
| Stochastic Gradient (SGD) | Joy, Fear, Sadness, Shame Guilt | 0.66 | 0.66 | 0.66 | 65.57 |
| Random Forest (RF 200) | Joy, Fear, Sadness, Shame Guilt | 0.64 | 0.64 | 0.64 | 64.02 |
| Naïve Bayes | Joy, Fear, Sadness, Shame Guilt | 0.64 | 0.64 | 0.64 | 66.58 |
| Logistic Regression | Joy, Fear, Sadness, Shame Guilt | 0.67 | 0.67 | 0.66 | 66.58 |
| K-Nearest Neighbor (KNN 25) | Joy, Fear, Sadness, Shame Guilt | 0.58 | 0.58 | 0.57 | 57.81 |
| Back Propagation Neural Classifier (BPN) | Joy, Fear, Sadness, Shame Guilt | 0.67 | 0.66 | 0.67 | 71.27 |

*Recommendation*

On the basis of results presented in Table 12, it is recommended that Back Propagation Neural Classifier (BPN) and logistic regression classifier have produced best results for the detection of different emotion categories (Joy, Fear, Sadness, Shame, and Guilt) from the text.

*4.4 Comparison with similar studies*

We evaluated the performance of the "Logistic Regression" Classifier, which exhibited better results in this work, with other similar studies conducted on for emotion classification. Table 13 shows the performance evaluation results. It is clear that the Logistic Regression (proposed work) performed better than the similar studies methods in terms of different evaluation measures such as accuracy, precision, recall, and f-score.

**Table 13:** Comparison with similar studies

| Study Reference | Methods/Techniques and Datasets | Experimental Results | | | |
|---|---|---|---|---|---|
| | | Accuracy (%) | Precision (%) | Recall (%) | F-Score (%) |
| Danisman, T & Alpkocak, A. (2008) | Classifier: Naïve Bayes<br><br>Dataset: ISEAR | 65.1 | 57 | 59 | 59 |
| Thomas et al. (2014) | Classifier: Naïve Bayes<br><br>Dataset: ISEAR | 64.23 | 61 | 62 | 62 |
| Our Work | 7 M.L classifiers<br><br>Classifier with Best Results: Logistic Regression<br>Dataset: ISEAR | 67 | 67 | 66 | 66.5 |

**5 Conclusions and Future Work**

This study performed a comparative analysis of performance evaluation of different machine learning classifiers for emotion detection and classification. This study is comprised of the following modules: i) data collection using 'ISEAR' dataset; ii) applying pre-processing on the acquired dataset, iii) applying machine learning classifiers for emotion detection, iv) comparison of different classifiers on the basis of their results, and v) recommending the best classifier for detection of emotion.

The proposed technique assists in classifying the text in different emotions like joy, fear, sadness shame, and guilt by using different machine learning classifiers: Random forest, *SVM, Logistic regression, Xgboost, SGD classifier, Naïve Bayesian, KNN*). The input text is categorized into different emotion categories like joy, fear, sadness, shame, and guilt. The performance of different classifiers is evaluated on the basis of their results. The experimental results in terms of precision, recall, f-measure, and accuracy, show that the Logistic Regression classifier outperformed other classifiers in terms of improved recall (83%), BPN yielded improved accuracy (71.27%) whereas the SVM yielded better results w.r.t precision (76%) and f-score (77%). As far as the worst-case analysis is concerned, the XGBoost has shown poor performance in terms degraded precision (66%), recall (66%), F-measure (66%) and accuracy (58.5%).

### *5.1 Limitations*
1. In the proposed work, experimentation is performed with five emotions categories, namely *joy, fear, sadness, shame, and guilt*. However, a further combination of emotions has not experimented.
2. Only one data set ("ISEAR") is used in the experiments with a subsample of five thousand records.
3. The random splitting technique is used in the experiment to split the data into training and testing.
4. Experiments are performed with respect to emotion detection on the classical machine learning classifiers namely SVM, Random forest, XGboost, KNN, Logistic regression, SGD classifier, and Naïve Bayesian.
5. Traditional features selection techniques are used in the experiments, such as TF-IDF and TF-IDF, which need to be replaced.

### *5.2 Future Directions*
1. To obtain more robust results, further experimentation is required with a different combination of emotions, such as fear_disgust, anger_disgust, and shame_guilt.
2. Multiple benchmark datasets of emotions detection, such as SemEval, and others can be used for carrying out the performance evaluation of the different machine learning classifiers.
3. In addition to random splitting, other techniques such as cross-validation can be applied for achieving more promising results.
4. Further experiments are required for emotion detection using deep learning techniques.
5. Instead of classical feature set used in the machine learning classifiers, automatic feature engineering, based on word embedding, can be investigated for emotion detection.

**Acknowledgement**
This publication has been supported by the Project: "Support of research and development activities of the J. Selye University in the field of Digital Slovakia and creative industry" of the Research & Innovation Operational Programme (ITMS code: NFP313010T504) co-funded by the European Regional Development Fund.

**Conflict of Interest:** All authors declare that they have no conflict of interest

**Ethical approval:** This article does not contain any studies with human participants performed by any of the authors and does not contain any studies with animals performed by any of the authors.